# Sr flux stability against oxidation in oxide-MBE environment: flux, geometry, and pressure dependence

Y.S. Kim[1], Namrata Bansal[2], Carlos Chaparro[1], Heiko Gross[1], and Seongshik Oh[1, a]

[1] Department of Physics & Astronomy, Rutgers, The State University of New Jersey, 136 Frelinghuysen Rd, Piscataway, NJ, 08854, U.S.A.

[2] Department of Electrical and Computer Engineering, Rutgers, The State University of New Jersey, 94 Brett Rd, Piscataway, NJ, USA

a) Electronic mail: ohsean@physics.rutgers.edu

MATERIAL NAMES; Strontium (Sr), Strontium Oxide (SrO), Oxygen ($O_2$)

# Abstract

Maintaining stable fluxes for multiple source elements is a challenging task when the source materials have significantly different oxygen affinities in a complex-oxide molecular-beam-epitaxy (MBE) environment. Considering that Sr is one of the most easily oxidized and widely used element in various complex oxides, we took Sr as a probe to investigate the flux stability problem in a number of different conditions. Source oxidation was less for higher flux, extended port geometry, and un-melted source shape. The extended port geometry also eliminated the flux transient after opening a source shutter as observed in the standard port. We also found that the source oxidation occurred more easily on the crucible wall than on the surface of the source material. Atomic oxygen, in spite of its stronger oxidation effectiveness, did not make any difference in source oxidation as compared to molecular oxygen in this geometry. Our results may provide a guide for solutions to the source oxidation problem in oxide-MBE system.

# I. INTRODUCTION

An oxidizing ambient is necessary for oxide-MBE growth, but it leads to several complications like source flux instability[1]. In particular, when multiple source materials with significantly different oxygen affinities are used, as in cuprate superconductors and multi-elemental transition metal oxides, maintaining stable fluxes for all elements is a challenging task[2-3]. Many of these oxides are composed of both alkaline-earth elements such as Sr and transition metal elements such as Cu. While Sr can fully oxidize even in ~$10^{-8}$ Torr of molecular oxygen[4], Cu requires much stronger oxidation conditions such as up to ~$10^{-5}$ Torr of ozone or atomic oxygen[2-3, 5-7]. However, if the alkaline-earth elements are exposed to such a strong oxidation condition, they tend to oxidize so much[1] that it is very difficult to control their fluxes better than 1% without a real time flux monitoring scheme such as atomic absorption spectroscopy[8-9], which significantly complicates the oxide growth. Although this source oxidation problem has been known to the complex-oxide MBE community since early 1990s[1], detailed studies are still lacking.

In this paper, we report how Sr flux changes due to source oxidation in different source configurations: Sr is the primary alkaline earth element in various complex oxides such as $SrTiO_3$, $La_{1-x}Sr_xMnO_3$, $La_{2-x}Sr_xCuO_4$, etc. Our objective through this study is to find out the optimal source conditions not only for Sr but also for other elements to achieve less than 1% flux variation over several hours of growth even in very harsh oxidation conditions.

## II. EXPERIMENT

We performed these experiments in custom-designed SVTA MOS-V-2 MBE system. The source port was modified from SVTA's standard design to allow two source-to-substrate distances: 21 cm for the standard and 42 cm for the extended port geometry as measured between the substrate to the orifice of the source crucible [Fig. 1]. The effusion cell axis was at an angle of 33° to the substrate normal. 40 cc pyrolytic boron nitride (PBN) and graphite crucibles were charged with 99.99% strontium sources (Aldrich-APL): one of the Sr sources was pre-melted to have a smooth top surface (service provided by Aldrich-APL) while the other was composed of random-shaped pieces and will be called un-melted from now on. Pre-melting allowed charging up to 75 g of Sr, while in the un-melted case, 15 g was almost the maximum. Low-temperature effusion cells (SVTA-275/450/458-XX) controlled by Eurotherm 2408 temperature controllers were used for thermal evaporation of sources. Stability of the source temperature was better than 0.1 °C, and flux drift was less than 1% over several hours when no oxygen was introduced, except right after the sources were charged. A RF plasma source (SVTA RF-4.5) with dissociation efficiency of up to 70%, RF frequency of 13.56 MHz, and 350 W of RF power[10] was used to generate the atomic oxygen. We used a differentially-pumped mass flow controller in combination with a precision leak valve to control the oxygen partial pressure from $10^{-9}$ to $10^{-4}$ Torr, while the system base pressure was ~$10^{-10}$ Torr during the experiments.

We measured the Sr flux using a quartz crystal microbalance (QCM: Inficon BDS-250, XTC/3): resonance frequency of 6 MHz, sampling rate of 4 Hz with 6.25 second rolling average, and measurement resolution of 0.0021 Å/sec [Fig. 2]. The QCM was mounted at the growth position, 21 cm apart from the orifice of the source crucible [Fig. 1], and water-cooled at a temperature of 20.0 °C. We also used atomic absorption spectroscopy (AA: SVTA AccuFlux-01/3)[8-9, 11] (chopping frequency of 511.2 Hz) to measure the Sr flux during short term flux-stability measurements. The light beam traversed underneath the substrate level through the vapor flux twice as it got reflected by a mirror on the opposite side of the vacuum chamber. Compared to QCM, one advantage of AA measurement is that it does not suffer from its own thermal transient effect when a source shutter is opened, and thus it is a better technique for separating out thermal transients of the source itself due to the shuttering. Because the measured flux signals are affected not only by source oxidation but also by scattering effect, we investigated the flux stability both in oxygen and in argon environment and compared their difference.

QCM signal was strongly affected by heat load from the source during opening and closing of the shutter. This is because the resonance frequency of the crystal can change not only by deposition of materials, which is the intended operation mode of QCM, but also by change of the crystal temperature. QCM signal jumps right after opening or closing of the source shutter and then decays until the crystal sensor reaches an equilibrium temperature [Fig. 2]. In order to take this into account, we excluded the first one hour data for long-term stability analysis. In addition, five minute moving average was used for all long-term results to improve the resolution and signal-to-noise ratio [Fig. 2].

## III. RESULTS AND DISCUSSION

Figure 3 shows the effect of source temperature, source-to-substrate distance, and source type (un-melted and pre-melted) on flux, measured by QCM. The flux depends exponentially on cell temperatures. The extended port, having twice the distance between the source and the substrate as compared to the standard port, provided four times less flux value. The larger effective surface area of the un-melted source as compared to the pre-melted one contributed to an increase in flux. The flux stability of all these configurations was studied under various oxygen conditions.

Figures 4 (a) and (b) show the effect of flux scattering in an argon environment. The source-to-substrate distance was a key factor in determining the extent of the flux scattering from the surrounding gas -- the longer the distance, the more the scattering. This process can be well described by the Beer-Lambert law. The transmission probability of a beam of flux through a gas environment can be written as:

$$\frac{I}{I_0} = e^{-\sigma n L} = e^{-\frac{L}{\lambda}} = e^{-\frac{\sqrt{2}\pi L d^2 P}{k_B T}}, \qquad (1)$$

where $I_0$ and $I$ are the intensities of the flux at the base pressure and the measurement pressure, respectively. $\sigma$ is the effective cross sectional area of the colliding gas species, $n$ is the number of gas particles per unit volume, $\lambda$ is the mean free path[12], $L$ is the distance between the source and the substrate, $k_B$ is the Boltzmann constant, $T$ is the temperature, and $d$ is the diameter of the gas particles. The result of this theoretical model, with $d = (r_{Ar} + r_{Sr}) = 3.1$ Å and $T = 300$ K, is consistent with the measured QCM-flux data in Ar environment [Fig. 4 (a) and (b)]. Even in the absence of any source oxidation, the measured short-term flux stability curves (the pressure dependence) shown in Fig. 4 (a) and (b) could never be higher than the Ar curve from the same source geometry because beam scattering would still exist regardless of source oxidation. In other words, the difference between each oxygen curve and the corresponding Ar curve should be considered the source oxidation effect.

Interestingly, QCM and AA provided quite different flux values at higher gas pressures: AA tends to overestimate the flux as compared to the value given by QCM. The crystal sensor of QCM is mounted at the growth position and the flux collecting area (diameter 0.8 cm) is similar to that of a standard substrate. Thus, QCM provides the actual amount of Sr being deposited on the substrate. But AA detects all Sr atoms present in the path of the light even those outside the substrate area. If the atomic beam coming from the source broadens due to scattering by gas molecules, it can still contribute to the AA flux signal, thus resulting in larger value than what QCM would provide. The magnitude of this overestimated flux value in AA depends on the details of geometrical alignment between the light and the source beam. This suggests that care should be taken in interpreting AA signal as proportional to the beam flux. We then introduced molecular oxygen to study short-term and long-term flux stabilities in an oxidizing environment for both the standard and the extended ports. For the short-term stability, oxygen pressure was increased from $1 \times 10^{-7}$ to $5 \times 10^{-5}$ Torr in steps and the corresponding flux was recorded for 1.5 min at each pressure. Lettieri *et. al* studied the degree of oxidation of Sr as a function of oxygen partial pressure and found that the Sr being deposited on the substrate, at a deposition flux of $7 \times 10^{13}$ Sr atoms/cm$^2$s (~ 0.4

Å/sec), starts to partially oxidize at pressures less than $3 \times 10^{-9}$ Torr and completely oxidizes at a pressure of $\sim 8 \times 10^{-8}$ Torr [4]. For most of our measurements, the Sr flux was less than this, so the Sr being deposited at the substrate was fully oxidized in our oxygen pressure range. The long-term source stability was tested by continuously monitoring the flux over four hours, keeping the oxygen pressure constant at $1 \times 10^{-5}$ Torr, a common pressure toward high side for oxide-MBE growth. The source shutter was opened after oxygen was introduced into the chamber. We started recording the data an hour after opening the source shutter, allowing the crystal sensor to reach its thermal equilibrium.

As shown in Fig. 4, Sr flux decreases on increasing the oxygen pressure. It can be explained by the oxide layer formed at the source surface. The vapor pressure of the oxide (SrO) is negligible at temperatures below 1600 K [13], so once the Sr source becomes oxidized, its vapor pressure starts to drop. This is also consistent with a previous report by Hellman and Hartford, who found that the flux of Mg, Ca, and Sr decreases exponentially with increasing oxygen pressure because of source oxidation[1]. But in case of Ba, its flux increases linearly with increasing oxygen pressure, implying that Ba surface oxide is unstable unlike the other alkaline earth oxides. In either case, source oxidation for alkaline earth elements results in significant changes in their flux, and thus has to be minimized in order to maintain stable fluxes. Irrespective of the geometrical configuration and source type, higher flux resulted in superior short-term and long-term stabilities. For the long-term stability, when the flux was higher than 0.3 Å/sec, the flux drift was less than 1%, our target value [Fig. 5(a)]. If we compare the cases with similar flux values, the un-melted and the pre-melted showed similar long-term stabilities [Fig. 5] even if the former was a little better in the short-term stability as shown in Fig. 4(c).

The extended port performed better in both short-term and long-term stabilities than the standard port. In the long-term test with a flux of ~0.1 Å/sec, the extended port suffered from only 1.5% drop in flux, compared to 2.5% in the standard port [Fig. 5]. A number of factors are responsible for this difference. First, the larger source-to-substrate distance for the extended port requires a higher vapor pressure at the source level to maintain a similar flux at the substrate, and this in turn contributes to reduced source oxidation at the source level. Second, unlike in the standard port, in the extended port Sr atoms are continuously deposited on the port wall of length 23 cm and diameter 6 cm in front of the source [Fig. 1] and they work as an effective oxygen getter. As a result, the oxygen partial pressure near the source should be lower for the extended port than for the standard port. Thus, the higher source vapor pressure in coalition with the lower oxygen pressure near the source makes the Sr flux more stable for the extended port as compared to the standard port.

Additionally, we investigated the difference in source oxidation between atomic and molecular oxygen with the extended port, at a flux of 0.12 Å/sec. Quite surprisingly, there was no difference between the two [Fig. 4(d)], although we initially expected that atomic oxygen would result in more noticeable source oxidation considering its stronger reactivity. This observation implies that oxygen atoms transform almost completely into molecular oxygen by the time they reach the Sr source, which is not in direct line-of-sight of the atomic source. We confirmed this scenario by a residual gas analyzer (RGA): when the RGA was located outside the line-of-sight of the atomic source, both molecular and plasma oxygen exhibited an identical spectrum; a line-of-sight measurement from a similar unit showed up to 70% dissociation efficiency[10]. This is in contrast to the other popular strong oxygen source, ozone, which does not completely transform to the molecular oxygen even after multiple scattering from the chamber walls according to our previous experience.

Furthermore, depth of the source inside the crucible, defined as the distance from the top surface of the source material to the crucible orifice, influenced the short-term flux stability. Fig. 6 shows the dependence of the short-term stability at comparable fluxes on the depth of the source. Overall, source oxidation occurred more significantly at larger depths. This phenomenon can be understood by the following scenario. Sr flux being deposited on the substrate (or QCM crystal) is composed of two contributions: one directly from the source surface and the other bounced from the crucible wall above the source surface. In other words, the crucible wall works as a secondary source with much larger surface area than the primary source area. As the depth of the source increases, the contribution from this secondary source becomes more pronounced. Large surface area of the secondary source increases the probability of Sr meeting the incoming oxygen molecules leading to easy oxidation of these Sr atoms. This explains why the oxygen-pressure-dependent Sr flux drop is more significant for larger source depths. It also implies that in an oxide-MBE environment, even regular source consumption will lead to a faster long-term flux drop than is expected from source consumption alone. Any solution for the source oxidation problem should properly handle this crucible wall issue as well.

Finally, we studied the flux transient effect related to source shuttering using atomic absorption spectroscopy (AA): AA is a better technique than QCM for this type of measurement because AA does not suffer from the thermal transient as QCM does.

Opening and closing the source shutter is the primary method to control the amount of source material deposited on a substrate in MBE. But on/off control of the shutter introduces flux transients[14-15], resulting in a poor control of composition when frequent shuttering is required. A closed shutter reflects the thermal radiation from the source back into the source, increasing the net source temperature. Due to this radiation and consequent rise in the temperature, the initial flux is always higher than the steady state value. To minimize such flux transient, Maki *et. al* used a conical insert in the source crucible[16] and Celii *et. al* modified the cell temperature before and after opening the shutter[17]. Our source shutter is designed to have an angle with the cell in order to reflect most of the radiation away from the source. Still, there was an observable flux transient associated with the standard port because of the short distance between the source and the shutter (4 cm). The initial flux right after shutter opening was 2% (0.003 Å/sec) higher than the steady state value (0.15 Å/sec) at a cell temperature of 440 ˚C [Fig. 7(a)]. The time constant of this flux transient was 1.3 min and it took ~2 min to reach the steady state after opening the shutter. The extended port did not exhibit such a flux transient because of the long shutter-to-cell distance (25 cm) [Fig. 7(b)].

## IV. CONCLUSIONS

We investigated the flux stability of Sr due to source oxidation in a number of different conditions and observed less source oxidation in higher flux, extended port geometry, and un-melted source shape. The extended port geometry also eliminated the flux transient as observed in the standard port. Source oxidation occurred more easily on the crucible wall than on the surface of the source material. There was not any observable difference between atomic and molecular oxygen in terms of source oxidation. We believe that this work will help finding solutions for the source oxidation problem in oxide MBE.

## ACKNOWLEDGMENTS

This work is supported by IAMDN of Rutgers University, National Science Foundation (NSF DMR-0845464) and Office of Naval Research (ONR N000140910749).

## Figure Captions

**Fig. 1.** Schematic diagram of the oxide-MBE system. The source port allows two different source-to-substrate distances: 21 and 42 cm.

**Fig. 2.** Raw and moving-averaged QCM data vs. time. PM stands for pre-melted. Five minute moving average was used to improve the resolution and signal-to-noise ratio of the QCM. Thermal heating of the crystal sensor created an erroneous transient signal right after the source shutter was opened. The exponential fit provided a time constant of 20 minutes. In order to get around this transient effect, we excluded the first one hour data for every long-term analysis.

**Fig. 3.** Sr flux (measured without oxygen) vs. temperature for all configurations. PM stands for pre-melted and UM un-melted. The effects of source temperature, source-to-substrate distance, and source type (un-melted and pre-melted) on flux can be seen.

**Fig. 4.** Short-term flux stability in argon and oxygen environment for (a) standard port, (b) extended port, and (c) similar flux conditions: the flux values in the legend were measured at base pressures. Pressure dependence of the argon curves is completely due to beam scattering, and source oxidation accounts for the difference between the oxygen curves and each argon curve.  Higher flux resulted in superior short-term stabilities.  AA tends to overestimate the flux as compared to QCM at higher gas pressures.  (d) Comparison of atomic and molecular oxygen: all other conditions were the same.  There is no observable difference between the two.

**Fig. 5.**  Long-term flux stability for (a) standard port and (b) extended port.  The extended port showed better stability than the standard port at comparable fluxes.

**Fig. 6.**  Short-term flux stability at comparable fluxes for different depths of the source.  Source oxidation became more pronounced with increased source depth, suggesting that there is preferential oxidation on the crucible wall of the source.

**Fig. 7.**  Flux transient at (a) standard port and (b) extended port.  The standard port showed ~2% shuttering transient, whereas the extended port did not show any observable transient.

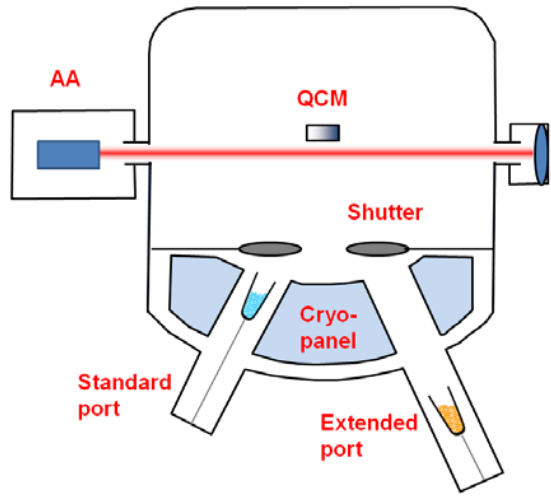

**Fig. 1**

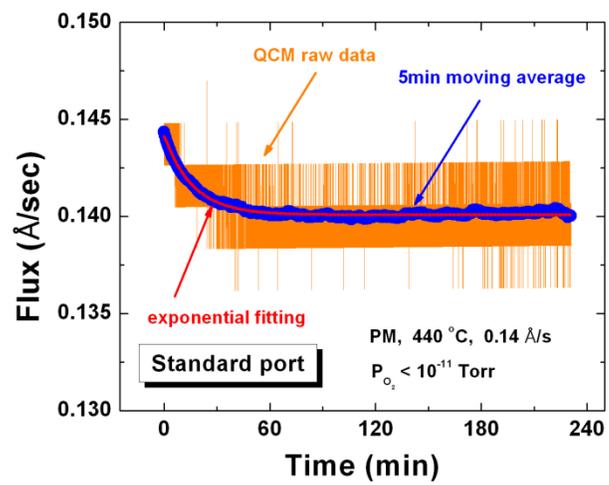

**Fig. 2**

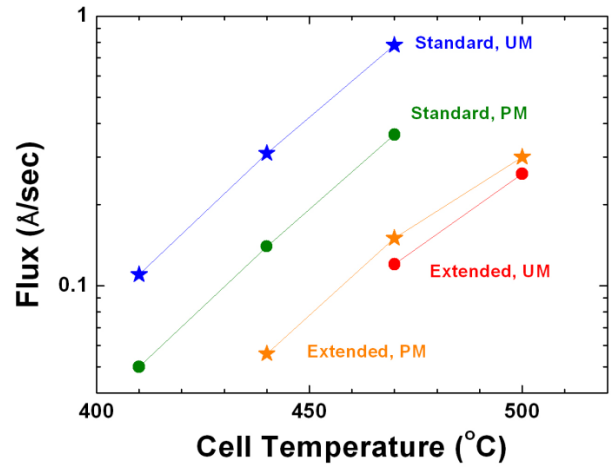

**Fig. 3**

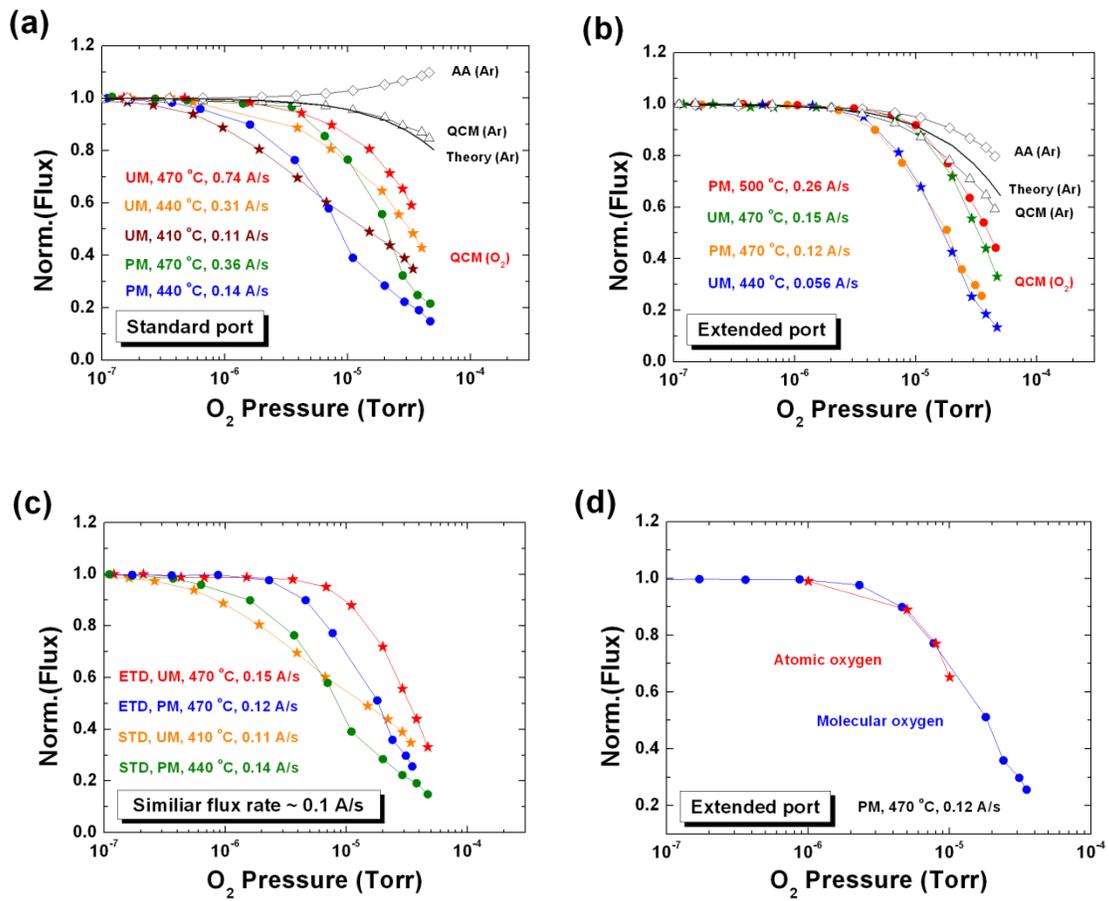

Fig. 4

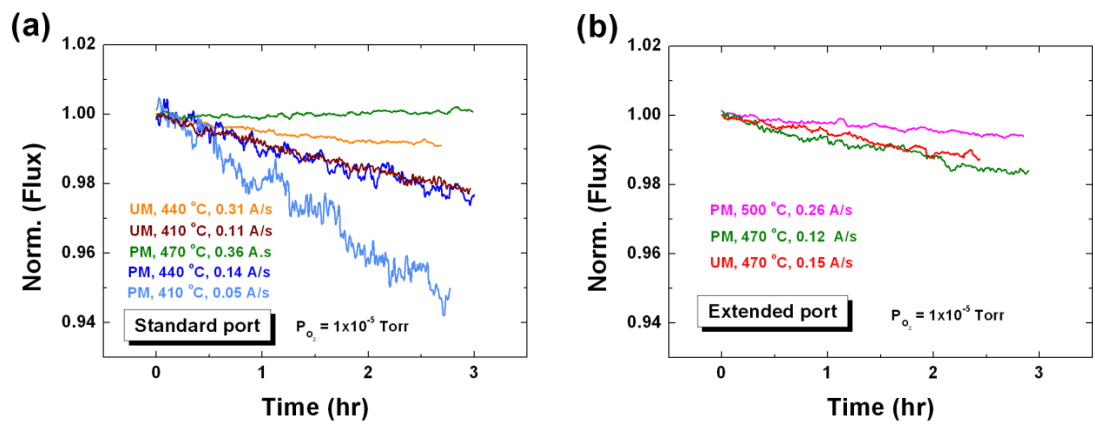

**Fig. 5**

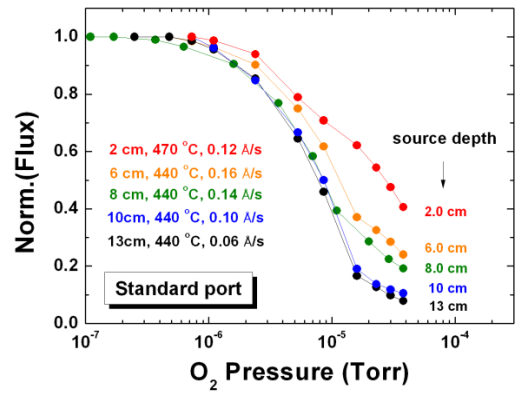

Fig. 6

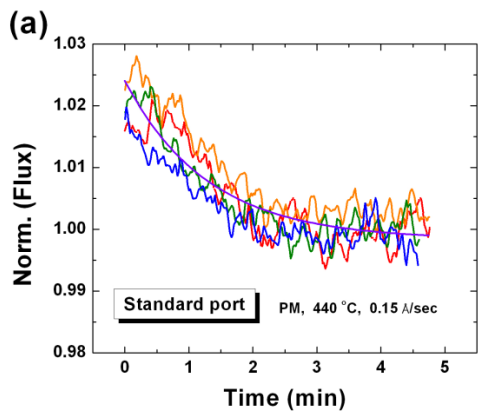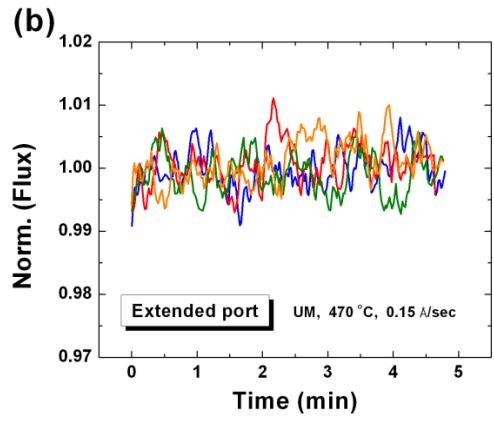

**Fig. 7**